%% file: main.tex
\pdfoutput=1

\documentclass[11pt]{article}

\usepackage[preprint]{acl}

\usepackage{times}
\usepackage{latexsym}

\usepackage[T1]{fontenc}

\usepackage[utf8]{inputenc}

\usepackage{microtype}

\usepackage{inconsolata}

\usepackage{graphicx}
\usepackage{enumerate}
\usepackage{booktabs}
\usepackage{enumitem}

\usepackage{amsfonts}       
\usepackage{nicefrac}       
\usepackage{multirow, multicol}
\usepackage[title]{appendix}
\usepackage[hang,flushmargin]{footmisc}
\usepackage[listings, skins]{tcolorbox}
\usepackage{pifont}
\usepackage{wrapfig}
\usepackage{caption}
\usepackage{subcaption}
\usepackage{arydshln}
\definecolor{darkred}{rgb}{0.5, 0.0, 0.0}
\definecolor{darkgreen}{rgb}{0.0, 0.5, 0.0}
\definecolor{whitesmoke}{rgb}{0.96, 0.96, 0.96}
\newtcblisting{code}{
  arc=0pt,
  outer arc=0pt,
  left=1mm,    
  right=1mm,   
  top=-1mm,     
  bottom=-1mm,  
  listing only,
  listing options={
    language=Python,
    basicstyle=\ttfamily\footnotesize,
    keywordstyle=\color{Blue},
    stringstyle=\color{darkred},
    numberstyle=\color{darkgreen},
    commentstyle=\color{darkgreen},
    breaklines=true,
    breakindent=0pt,
    showstringspaces=false,
    lineskip=-1pt,
  },
  boxrule=0.5pt,
  colback=whitesmoke,
  colframe=black,
}
\usepackage[T1]{fontenc}
\usepackage{ascii}
\usepackage{scrextend}

\newenvironment{prompt}{%
  \begin{tcolorbox}[
    arc=0pt,
    outer arc=0pt,
    boxrule=0.8pt,
    colback=whitesmoke,
    fontupper=\ttfamily,
    colframe=gray!50!black,
    sharp corners=southwest,
    enhanced
  ]%
}{%
  \end{tcolorbox}%
}

\renewcommand{\textcolor}[2][]{\ignorespaces}

\newcommand{\target}{\sc{target}}

\title{TARGET: Benchmarking Table Retrieval for Generative Tasks}

\author{\textbf{Xingyu Ji\thanks{Correspondence to \href{mailto:madelon@cwi.nl}{madelon.hulsebos@cwi.nl} and \href{mailto:jixy2012@berkeley.edu}{jixy2012@berkeley.edu}}\textsuperscript{1}}, \textbf{Parker Glenn\textsuperscript{2}}, \textbf{Aditya G. Parameswaran\textsuperscript{1}}, \textbf{Madelon Hulsebos\textsuperscript{3}}
\\
 \textsuperscript{1}UC Berkeley \ \textsuperscript{2}Capital One \ \textsuperscript{3}CWI
}

\begin{document}
\maketitle

\input{sections/0_abstract.tex}

\input{sections/1_introduction.tex}

\input{sections/2_related_work.tex}

\input{sections/3_target.tex}

\input{sections/4_results.tex}

\input{sections/5_conclusion.tex}

\bibliography{custom}

\appendix

\input{sections/6_appendix.tex}

\end{document}

%% file: sections/0_abstract.tex
\begin{abstract}
The data landscape is rich with structured data, often of high value to organizations, driving important applications in data analysis and machine learning. Recent progress in representation learning and generative models for such data has led to the development of natural language interfaces to structured data, including those leveraging text-to-SQL. Contextualizing interactions, either through conversational interfaces or agentic components, in structured data through retrieval-augmented generation can provide substantial benefits in the form of freshness, accuracy, and comprehensiveness of answers. The key question is: how do we retrieve the right table(s) for the analytical query or task at hand? To this end, we introduce {\target}: a benchmark for evaluating \textbf{TA}ble \textbf{R}etrieval for \textbf{GE}nerative \textbf{T}asks. With {\target} we analyze the retrieval performance of different retrievers in isolation, as well as their impact on downstream tasks. We find that dense embedding-based retrievers far outperform a BM25 baseline which is less effective than it is for retrieval over unstructured text. We also surface the sensitivity of retrievers across various metadata (e.g., missing table titles), and demonstrate a stark variation of retrieval performance across datasets and tasks. 
{\target} is available at 
\texttt{\url{https://target-benchmark.github.io}}.

\end{abstract}

%% file: sections/1_introduction.tex
\section{Introduction}\label{sec:introduction}

Large Language Models (LLMs) have  become an indispensable tool in the knowledge worker's arsenal, providing a treasure trove of information at one's fingertips. Retrieval-Augmented Generation (RAG)~\cite{lewis2020rag} further extends the capabilities of these LLMs by grounding generic dialog using information from external data stores. Despite progress in long-context LLMs, RAG still provides benefits in cost and inference time~\cite{li2024longcontextrag, yu2024defenseofrag}. Moreover, it allows us to augment generic, off-the-shelf LLMs with proprietary data they haven't been trained on. 
Progress on RAG has largely been enabled by benchmarks that help exhaustively evaluate the effectiveness of various methods~\cite{yang2024crag, muennighoff2023mteb}.

\begin{figure*}
    \centering
    \includegraphics[width=1\textwidth]
    {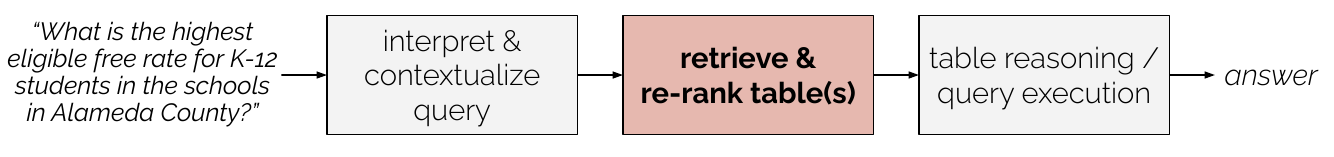}
    \caption{Pipeline of ``open domain'' question answering over tabular data, in which no tables containing ``evidence'' for the question are provided. Most research considers the ``closed domain'' setting, and focuses on query interpretation and augmentation, or table reasoning or query generation (e.g. text-to-SQL). With {\target} we intend to stimulate and facilitate research on the critical retrieval step in the ``open domain'' setting.}\label{fig:e2e-pipeline}
\end{figure*}

While RAG has been extensively explored for free-form text, this is unfortunately not the case for structured data, stored either in relational databases or otherwise. Prior work has shown that structured data is of a different nature, for example regarding data types and dimensionality, requiring dedicated research~\cite{cong2023observatory}. Moreover investigating retrieval of structured data for RAG is important: contextualizing LLMs using frequently updated statistical data sources, such as Data Commons~\cite{guha2023datacommons}, or using proprietary relational databases within organizations, can yield rich dividends~\cite{radhakrishnan2024datacommonsrag}, all underscoring the need for better models, approaches and evaluation for retrieval over structured data.

Another important motivation for research on table retrieval stems from research on LLM-powered interfaces and agentic systems for processing and querying structured data. Most research in this direction, e.g., for question answering~\cite{nan2022fetaqa} or text-to-SQL~\cite{gao2023dailsql}, assumes that a table or relational database is provided, while identifying the relevant table is, in fact, a non-trivial task for a user (or agent). Figure~\ref{fig:e2e-pipeline} depicts an end-to-end pipeline as we envision: starting with a natural language query (which can be a ``lookup'' or analytical question), the first step is to interpret and augment the query, for which the retrieval component identifies the relevant tabular data needed to generate a response (which can be in code, natural language, or other format). We find that table retrieval in end-to-end (analytical) query systems is an understudied area, motivating the creation of a benchmark.

While there has been initial work exploring open-domain question answering on public table corpora such as Wikipedia~\cite{chen2020ottqa, herzig2021dtr}, this does not represent the full spectrum of data characteristics and tasks for structured data retrieval. The development of a broad and comprehensive benchmark covering diverse tasks and datasets of varying difficulty is therefore key in advancing retrieval systems for structured data.

In this paper, we present {\target}: {\em the first benchmark evaluating Table Retrieval for Generative Tasks}. With {\target} we provide a consistent and comprehensive framework for evaluating models and pipelines for table retrieval in isolation, as well as end-to-end for downstream tasks. We use {\target} to analyze retrieval methods based on sparse lexical representations~\cite{chen2020ottqa}, dense embeddings of metadata~\cite{Liu_LlamaIndex_2022},  dense table embeddings~\cite{zhang2025jasperstelladistillationsota}, and dense row embeddings~\cite{kumar2023mitqa}. We find that sparse lexical representations are far less effective for retrieval over tabular data as it is found to be for rich free-form text~\cite{muennighoff2023mteb}. In our analysis with {\target}, we find that dense table- and row- embeddings~\cite{zhang2025jasperstelladistillationsota} outperform baselines but still show high variation in performance across tasks and datasets. Finally, we highlight the sensitivity of retrievers to the provided metadata inputs (e.g., web page titles) and table data availability (e.g., embedding full tables, column names only, or generated table summaries). Our findings identify a performance gap in retrieval accuracy and robustness across data and tasks, emphasizing the need for more research in this area for which {\target} is an instrumental stepping stone.

%% file: sections/2_related_work.tex
\section{Related Work}\label{sec:related-work}

\begin{figure*}[ht!]
	\centering
	\includegraphics[width=\textwidth]
	{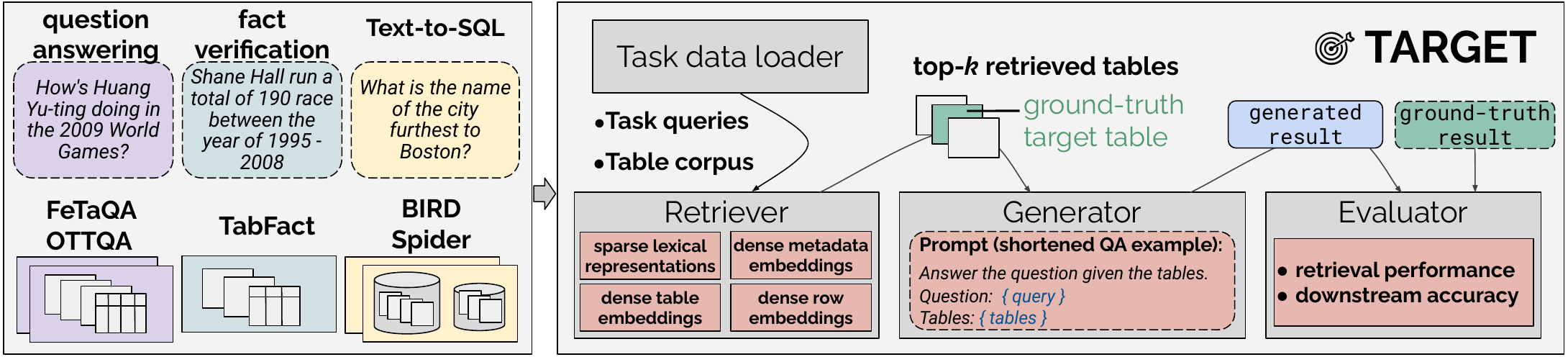}
	\caption{Overview of the {\target} benchmark for evaluating table retrieval methods and downstream generation for various datasets across three downstream tasks: tabular question answering, fact verification, and Text-to-SQL.}\label{fig:target_diagram}
    \vspace{-0.2cm}
\end{figure*}

\paragraph{Representation Learning and LLMs for Tables} Tables have recently become a modality of interest for representation learning and generative models for tasks such as table understanding~\cite{hulsebos2019sherlock, deng2022turl}, fact verification~\cite{herzig2020tapas, zhang2020tablefact}, and question answering~\cite{herzig2020tapas}, and more recently text-to-SQL~\cite{gao2023dailsql}. These models either deploy LLMs out-of-the-box for tabular data, or develop tailored architectures to capture the properties of tables, which pose specific challenges~\cite{cong2023observatory}. These models typically take one or more tables and a query as input to generate an answer, however, the relevant tables per query can be difficult to identify. {\target} is intended to close this gap and facilitate research on end-to-end querying over tabular data such as text-to-SQL and question answering.

\paragraph{Table Retrieval} Retrieval of structured data has been studied across use-cases in data management and machine learning. Dataset search where the objective is to find a dataset for a given task (e.g. training a machine learning model or doin data analysis) is a well studied topic in the data management literature~\cite{halevy2016goods, castelo2021auctus}. These table retrieval systems typically take a semantic description of the data as input and return the relevant tables. In {\target} we focus on retrieval components embedded into end-to-end query systems, where input queries are natural language queries and the task is to provide an accurate response based on relevant data that first needs to be retrieved in an end-to-end manner. Such pipelines have mainly been studied for open-domain question answering, typically over web table corpora~\cite{chen2020ottqa, herzig2021dtr, wang2023solo}. We include OTTQA~\cite{chen2020ottqa}, a sparse lexical retriever, as a baseline for open-domain QA. We also integrate two commonly used datasets for open domain table QA (FeTaQA~\cite{nan2022fetaqa} and OTTQA~\cite{chen2020ottqa}) into {\target}. We introduce two new end-to-end query tasks: fact verification and text-to-SQL, which are typically not considered in the ``open-domain'' setting but assume the relevant data is provided by a user.

\paragraph{Benchmarks and Datasets} To develop stronger rerievers and advance research on LLM-driven tasks on structured data, benchmarks and datasets are essential. The MTEB and CRAG benchmarks~\cite{muennighoff2023mteb, yang2024crag} have been instrumental in benchmarking text embedding quality and RAG over rich text documents. We need similar benchmarks for retrieval systems and embedding models for structured data. In prior research, useful datasets were introduced to evaluate various tasks for relational data, such as TabFact~\cite{chen2019tabfact}, FeTaQA~\cite{nan2022fetaqa}, and Spider~\cite{yu2018spider}. These datasets focus on evaluating methods for a specific downstream task only, i.e., given a table or database, answer natural language queries about it, without integrating the critical task of retrieval. {\target} addresses this gap by focusing on the evaluation of table retrieval performance while incorporating existing task-specific datasets.

%% file: sections/3_target.tex
\section{The TARGET Benchmark}\label{sec:target}

We describe the datasets, tasks, metrics, and retrievers that make up the {\target} benchmark. All resources for use and extension of {\target} are available at 
\url{https://anonymous.4open.science/r/target-B782}{\footnote{The URL of the project page with links to datasets and code will be linked upon publication.}.

\subsection{Benchmark Design} The pipeline of {\target} aligns with typical RAG pipelines (Figure~\ref{fig:target_diagram}). {\target} takes as inputs the corpus with tables/databases and queries (a natural language question or statement). Data loading and evaluation are abstracted away such that custom core components of RAG pipelines, i.e., the \texttt{Retriever} and \texttt{Generator} can easily be evaluated when aligned with the {\target} API. The retriever, which can be basic or advanced~\cite{gao2023ragsurvey}, identifies the relevant table(s)/database(s) for an input query. Depending on needs, retrievers can either manage corpus embedding independently or leverage vector databases \cite{malkov2018hnsw, qdrant} integrated in {\target}. Given the tables and query, the generator yields a response which is then evaluated with respect to the ground-truth.

\subsection{Tasks \& Metrics} Per source dataset, we combine all tables and any available metadata into a retrieval corpus. For all tasks, e.g., question answering, we evaluate the retriever and generator outputs using metrics from the original dataset papers or that are widely adopted. An overview of the tasks and metrics in {\target} can be found in Table~\ref{table:tasks_datasets_metrics}.

\begin{table}[t!]
    \centering
    \caption{Tasks and Evaluation Metrics in {\target}.}
    \vspace{-0.2cm}
    \label{table:tasks_datasets_metrics}
    \renewcommand{\arraystretch}{1.2}
    \resizebox{\columnwidth}{!}{%
        \begin{tabular}{lll}
            \toprule
            \textbf{Task} & \textbf{Evaluation Metrics} \\
            \midrule
            Table retrieval & recall (R@$k$), capped recall (CR@$k$)\\
            &  avg. retrieval time (s) \\
            Question answering & SacreBleu (SB))\\
            Fact verification & precision (P), recall (R), F1-score (F1) \\
            Text-to-SQL & execution accuracy (EX)\footnotemark[2]\\
            \bottomrule
        \end{tabular}
    }
    \vspace{-0.3cm}
\end{table}

\paragraph{Table Retrieval} Table retrieval task assesses retrieval performance in isolation and is the first step for end-to-end downstream evaluation. Retrieval performance is measured with recall@top-$k$, reflecting the successful retrieval of the ground-truth table within the top-$k$ retrieved tables. In the text-to-SQL setting, however, standard recall may yield unintuitive results, as multiple ground-truth tables might be needed to generate the valid SQL query. With $T_i$ representing the ground-truth tables for the $i$th query, we correct for situations where $k \ll |T_i|$ and follow \citet{thakur2021beir} in evaluating capped recall by setting our denominator to $min(k, |T_i|)$. Additionally, we include the average retrieval time per query.

\paragraph{Question Answering} Given the retrieved tables contents and the input question, an answer is generated and evaluated against the ground-truth natural language answer for accuracy and comprehensiveness. We report SacreBleu~\cite{post-2018-call} to reflect syntactic similarity across generated tokens. 

\paragraph{Fact Verification} Given the retrieved tables, the generator either accepts or refutes a natural language statement, or acknowledges that insufficient information is provided. Here, the accuracy is measured through precision, recall and F1.

\paragraph{Text-to-SQL} We adapt a prompt template from \citet{talaei2024chess}, which incorporates the natural language question and the schemas of the retrieved tables along with generation instructions. The prompt instructs the generator to output a concise ``chain-of-thought'' reasoning  trace~\cite{nye2021show, wei2022chain} to support more robust query generation. Additionally, since the retrieved tables may belong to different databases, the generator is required to include the selected database alongside the SQL query to ensure proper execution. The execution result from the generated SQL are then compared to that of the ground-truth SQL. We report the execution accuracy, aggregated across query complexity categories, following the implementation in BIRD \cite{li2024bird}\footnote{See \href{https://github.com/bird-bench/mini_dev/blob/47f116d828a45d139c0261cf804e610e031b572a/evaluation/evaluation_ex.py}{evaluation\_ex.py}}.

\subsection{Datasets}\label{app:datasets}

\paragraph{Data and Label Sources}

The datasets of each task in {\target} can be found in Table~\ref{table:dataset_specifications}. All publicly available splits of each dataset are included except for BIRD's train split. We use the test splits of included datasets for our evaluations. For OTTQA and BIRD, where test splits are unavailable, validation splits are used.

To ensure consistency across datasets, e.g. for consistent data processing, we standardize the schemas of the files holding the datasets. Each dataset has a ``corpus'' and a ``queries'' file. The ``corpus'' files contain the table contents and table identifiers (IDs), wherein each entry corresponds to a single table and includes a ``context'' field for metadata, if available. For instance, in the text-to-SQL datasets, the context field contains primary key, foreign keys, and other table schema information. The ``queries'' files contain the queries, query IDs, and the ground-truth table ID(s).

To evaluate retrieval for text-to-SQL, we extract all the tables referenced in the ground-truth query using \href{https://github.com/tobymao/sqlglot}{sqlglot} and consider them as ground-truth.

\paragraph{Data Complexity}
\textcolor{blue}{Tables across datasets differ significantly in size. For example, text-to-SQL datasets feature significantly larger tables compared to other datasets in {\target}. Although BIRD's validation split contains fewer tables and databases overall, the large size of each table poses a significant challenge for retrieval systems. Specifically, the average number of rows per table in BIRD is 52.4k, nearly 10x compared to 5.3k rows per table in Spider. In contrast, tables in FeTaQA, OTTQA, and TabFact range from 10 to 50 rows. The distributions of row and column counts per dataset can be found in Appendix~\ref{app:data-characteristics}.}

Another distinction across datasets is the availability of metadata. Unlike text-to-SQL datasets, which feature descriptive table names and database schema, FeTaQA and TabFact does not provide informative table titles (for example, ``2-1570274-4.html.csv'' from TabFact) or grouping by databases. This requires retrieval methods to effectively use tabular data contents or devise data augmentation methods.

\begin{table}
    \centering
    \caption{Dimensions of included tabular datasets per task across splits in {\target}.}
    \label{table:dataset_specifications}
    \renewcommand{\arraystretch}{1.2}
    \resizebox{\columnwidth}{!}{%
    \begin{tabular}{cllcc}
        \toprule
        \textbf{Task} & \textbf{Dataset} & \textbf{Split} & \textbf{Corpus size} & \textbf{\# queries} \\
        \midrule
        \multirow{5}{*}{\rotatebox[origin=c]{90}{\textbf{\small{Question Answering}}}} & OTTQA & train & 8.1K tables & 41.5K \\
                    &     &  val  & 789 tables & 2.2K  \\\cdashline{2-5}
                    & FeTaQA & train & 7.3K tables & 7.3K \\
                    &      & val  & 1K tables & 1K\\
                    &      & test & 2K tables & 2K\\\cdashline{2-5}
                    
        \multirow{3}{*}{\rotatebox[origin=c]{90}{\textbf{\small{Fact Verif.}}}} & TabFact & train & 13.2K tables & 92.3K \\
                    &      & val & 1.7K tables & 12.8k\\
                    &      & test & 1.7K tables & 12.8K\\\cdashline{2-5}
                    
        \multirow{4}{*}{\rotatebox[origin=c]{90}{\textbf{\small{Text-to-SQL}}}} & Spider & train & 146 DBs (2K tables) & 8.7K \\
                    &      & val  & 20 DBs (1K tables) & 1K\\
                    &      & test & 40 DBs (1K tables) & 2.1K\\\cdashline{2-5}
                 & BIRD & val & 11 DBs (75 tables) & 1.5K \\
        \bottomrule
    \end{tabular}
    }
\end{table}

\subsection{Retrievers}\label{sec:retrievers} We present our analysis with {\target} for retriever methods that reflect common design principles in research and industry. We evaluate dense semantic embeddings and sparse lexical representations, and vary the inputs provided: tables or rows, with or without table metadata, and metadata-only. Text-to-SQL has small changes to the retriever and generator, as explained in Appendix~\ref{app:prompts}.

\paragraph{No Context baseline} LLMs are capable of memorizing facts from the data that they were trained on~\cite{mallen2022llmmemory}. To understand the influence of memorization on downstream task responses, the LLM-based generator is asked to respond based soely on its internal knowledge without any retrieved tables provided. We refer to this setting as the ``No Context'' baseline.

\paragraph{Sparse Lexical Representation} The Sparse Lexical Representation retriever resembles the OTTQA approach~\cite{chen2020ottqa}. It constructs a TF-IDF matrix of the corpus , which may use TF-IDF term weights or BM25. It takes as input the column names, table rows, and, table metadata such as the (Wikipedia) page title. On retrieval, a query is converted into a TF-IDF-weighted vector for which the dot product is calculated with the table representations to find the $k$-most similar tables.

\paragraph{Dense Metadata Embedding}
While metadata such as titles and descriptions can provide context for retrieval, they are either uninformative (e.g. ``8c4c-4f0d.csv'') or entirely absent in many tables. To this end, the Dense Metadata Embedding retriever creates table summaries following three steps, \ding{172} generate a table name and summary of each table with GPT-4o-mini\footnote{gpt-4o-mini-2024-07-18\label{gpt}} using the column names and first 10 rows of the table, \ding{173} embed the table metadata with \texttt{text-embedding-ada-002}, and \ding{174} retrieve relevant tables based on the cosine similarity between natural language query and metadata embedding. We use the open-source LlamaIndex library, commonly used in practice, to store the embeddings in an in-memory key-value index and retrieve using cosine similarity \cite{Liu_LlamaIndex_2022}.

\textcolor{blue}{\paragraph{Dense Table Embedding} We compare three dense embedding models: \texttt{text-embedding-3-small} ~\cite{openai2024textembedding3small}, \texttt{stella\_en\_400M\_v5} ~\cite{zhang2025jasperstelladistillationsota}, and \texttt{multilingual-e5-large-instruct} ~\cite{wang2024multilinguale5textembeddings}\footnote{Experiments with the \texttt{tapas-large} model~\cite{herzig2020tapas} illustrated that this BERT-based table-specialized embedding model is not competitive as-is for retrieval.}. The latter two are open-weight models available on HuggingFace. We evaluate the performance for embeddings of only column names versus column names along with 100 rows. While formatting tables as json appeared better for GPT-3.5~\cite{singha2023tabular}, markdown formatting yields better results. Each row of the table is formatted in markdown's tabular syntax and sequentially appended to form a single concatenated string for embedding. For retrieval, the input query is embedded with the same model, and the top-$k$ tables are retrieved based on cosine similarity.}

\textcolor{blue}{\paragraph{Dense Row-level Embedding} The input query might semantically correspond to values of certain rows within tables. Alternative approaches, therefore, devise retrieval through row-level embeddings~\cite{zhang2023taptap,kumar2023mitqa,wang2023solo}. In this baseline, each row is serialized into a sentence following the template ``[column name]$_i$ is [cell value]$_i$, [column name]$_j$ is [cell value]$_j$''~\cite{zhang2023taptap}, for example, ``first name is John, last name is Doe''. The serialized rows are embedded using the relatively small and effective \texttt{stella\_en\_400M\_v5} embedding model (435M parameters). Upon retrieval, the input query is embedded with the same model and used for retrieving rows with the highest cosine-similarity with the input query embedding. Based on the retrieved rows, the corresponding top-$k$ tables are retrieved. Row-wise retrieval via dense embeddings can become impractical for very large tables with hundreds of thousands of rows, for example those included in BIRD. Therefore, this baseline is not evaluated for the BIRD dataset.}

\subsection{Generators}\label{sec:generators}  
We use basic LLM prompts for downstream tasks to evaluate the GPT-4o-mini model\footref{gpt} in our experiments \cite{hurst2024gpt}. However, we design the {\target} API to enable evaluations of other language models and advanced generation pipelines. 

The \texttt{Instruction} prompt takes in: \ding{172} task instructions, \ding{173} the top-$k$ retrieved table(s) or database schemas of retrieved tables (for text-to-SQL), and \ding{174} the query. Unless otherwise specified, we serialize all tables in prompts to markdown strings. An example prompt for the question answering task is provided below.  The full prompt templates can be found in Appendix~\ref{app:prompts}.

\begin{prompt}
\scriptsize{
Use the provided table(s) to answer the question. Yield a concise answer to the question.\\
If none of the tables provide relevant information, use your knowledge base to generate an answer — but only if you are confident in the answer's factuality.\\
If the neither the tables nor your knowledge can be used to answer the question reliably, say that not enough information is provided.\\
Tables: {\color{blue}\{table\_contents\}} \\
Question: {\color{blue}\{query\}}}
\end{prompt}

%% file: sections/4_results.tex
\section{Results}\label{sec:results}
Table~\ref{tab:retrieval-results} presents the performances of the evaluated retrievers with $k$ set to 10.  Figure~\ref{fig:top-k} illustrates the average retrieval recall over various values of $k$ across datasets. For the Sparse Lexical Retriever, only the performance using BM25 is included as its performance is similar to TF-IDF.

\subsection{Retrieval Insights}

\begin{table*}[h]
	\centering
    \textcolor{blue}{
    \caption{\textcolor{blue}{Results with {\target} for table retrieval with $k$=10. R@$k$ stands for recall@$k$, CR@$k$ stands for capped recall @$k$~\cite{thakur2021beir}, and s for average retrieval time in seconds. For the Dense Table Embedding baseline, we report the best performing model \texttt{stella\_en\_400M\_v5}. Best scores are in \textbf{bold}, second-best \underline{underlined}.}}\label{tab:retrieval-results}
	\renewcommand{\arraystretch}{1.2}
	\resizebox{\textwidth}{!}{%
		\begin{tabular}{l : c c c c  : c c  : c c c c }
			\toprule
			& \multicolumn{4}{: c :}{\textbf{Question Answering}} & \multicolumn{2}{c :}{\textbf{Fact Verification}} & \multicolumn{4}{c}{\textbf{Text-to-SQL}} \\
			& \multicolumn{2}{: c}{\textbf{OTTQA}}  &  \multicolumn{2}{c:}{\textbf{FeTaQA}} & \multicolumn{2}{c:}{\textbf{TabFact}} & \multicolumn{2}{c}{\textbf{Spider}} & \multicolumn{2}{c}{\textbf{BIRD}}\\
			\textbf{Method}  & R@10 & time (s)  & R@10 & time (s)  & R@10 & s  & CR@10 & time (s)  & CR@10  & time (s)  \\
			\midrule
			Sparse Lexical Repr. (BM25) & \textbf{0.967} & 0.001    & 0.082 & 0.001   & 0.338 & 0.001  & 0.544 & 0.001   & 0.700 & 0.001 \\
			\ \ \ \ \ \ \ \textit{w/o table title} & 0.592 & 0.001    & 0.084 & 0.001   & 0.331 & 0.001  & 0.491 & 0.001   & 0.616 & 0.001  \\
			Sparse Lexical Repr. (TF-IDF) & \underline{0.963} & 0.001 &  0.083   & 0.001   & 0.336 & 0.001  & 0.541 & 0.001   & 0.586 & 0.001 \\
			\ \ \ \ \ \ \ \textit{w/o table title} & 0.583 & 0.001 & 0.039   & 0.001   & 0.322 & 0.001  & 0.489 & 0.001   & 0.613 & 0.001 \\
			Dense Metadata Embedding & 0.820  & 0.297	 &	0.436 &	0.396 	& 0.469 & 0.354	 &	0.621 & 0.024 &	\underline{0.940} &	0.014	 \\
			Dense Table Embedding & \underline{0.963} & 0.001   & \textbf{0.741} & 0.001 & \underline{0.824} & 0.001  & \underline{0.657} & 0.001 & \textbf{0.961} &  0.003 \\
			\ \ \ \ \ \ \ \ \textit{column names only} & 0.658   & 0.001 & 0.208 & 0.001 & 0.506 & 0.001 & 0.648 & 0.001  & 0.932 & 0.003  \\
            Dense Row-level Embedding & 0.951 & 0.267 & \underline{0.711} & 0.394 & \textbf{0.848} & 0.384 & \textbf{0.665} & 6.077 & N/A & N/A \\
			\bottomrule
		\end{tabular}
	}
    }
\end{table*}

\begin{table}
  \centering
  \textcolor{blue}{
  \caption{\textcolor{blue}{Table Retrieval Performances of Dense Table Embedding with Various Text Embedding Models with $k$=10. Best scores are in \textbf{bold}, second-best \underline{underlined}.}}\label{tab:embedding-comparison-retrieval}
  \renewcommand{\arraystretch}{1.2}
  \resizebox{0.49\textwidth}{!}{%
    \begin{tabular}{l : c  c  : c  : c c }
      \toprule
      & \multicolumn{2}{: c :}{\textbf{Question Answering}} & \multicolumn{1}{c :}{\textbf{Fact Verif.}} & \multicolumn{2}{c}{\textbf{Text-to-SQL}} \\
      & \multicolumn{1}{: c}{\textbf{OTTQA}}  &  \multicolumn{1}{c:}{\textbf{FeTaQA}} & \multicolumn{1}{c:}{\textbf{TabFact}} & \multicolumn{1}{c}{\textbf{Spider}} & \multicolumn{1}{c}{\textbf{BIRD}}\\
      \textbf{Method}  & R@10  & R@10   & R@10  & CR@10  & CR@10  \\
      \midrule
       text-embedding-3-small  & \underline{0.950}   & \underline{0.722}   & \underline{0.779}   & 0.618 &  0.858  \\
      \ \ \ \ \ \ \ \textit{column names only}   &  0.601    & 0.184    & 0.452  & 0.635   & 0.908  \\
        stella\_en\_400M\_v5  & \textbf{0.963}   & \textbf{0.741}   & \textbf{0.824}  & \textbf{0.657}  & \textbf{0.961} \\
      \ \ \ \ \ \ \ \textit{column names only}  & 0.658   & 0.208   & 0.506  & \underline{0.648}  & \underline{0.932}  \\
      \small{multilingual-e5-large-instruct}   & 0.918   & 0.655   & 0.741   & 0.620   & 0.894 \\
      \ \ \ \ \ \ \ \textit{column names only}  & 0.549  & 0.188   & 0.430    & 0.613   & 0.909 \\
      \bottomrule
    \end{tabular}
  }
  }
\end{table}

\paragraph{How do different table representations perform?} We find that table retrieval based on sparse lexical representations such as BM25 (OTTQA) are less effective, across tasks and datasets, than they are for text~\cite{muennighoff2023mteb}, even with increased $k$ (Table ~\ref{tab:retrieval-results}). The strong performance of the sparse lexical retrievers with table title on the OTTQA dataset (recall@10 of 0.967 and 0.963) can be attributed to the high correspondence between Wikipedia table titles and the questions, as manually verified\footnote{The queries and tables can be explored at: \url{https://ott-qa.github.io/explore.html}}. The performance drops for BM25 and TF-IDF to 0.592 and 0.583 respectively, if the table title is not included. The importance of descriptive metadata for retrievers based on lexical representations is confirmed by their low performances on FeTaQA and TabFact, where descriptive table titles are not available. LLM-generated table summaries with dense metadata embeddings can significantly improve retrieval performance as illustrated by the Dense Metadata Embedding baseline. 

\textcolor{blue}{Dense Table Embeddings (with column names and rows included in the embeddings) generally yield the best performance. Different embedding models demonstrate similar performance across datasets, with \texttt{stella\_en\_400M\_v5} achieving the best results, showing itself to be a viable open-source, lightweight, and efficient option (Table \ref{tab:embedding-comparison-retrieval}). Notably, for both text-to-SQL datasets, the effect of including data rows is minimal, with differences within $\pm 5\%$ in recall. Inspection confirms that (analytical) queries in text-to-SQL datasets typically have high resemblance with schemas (column names). In contrast, for the question answering and fact verification tasks, retrieval performances are significantly curtailed when only the column names are embedded.}

\textcolor{blue}{The Dense Row-level Embedding method exhibits comparable performances to dense embeddings of tables with sampled rows. On Question Answering datasets, row-level retrieval does not improve performance compared to dense table embeddings, while for Text-to-SQL and Fact Verification it lightly outperforms other baselines. However, due to large size tables for Text-to-SQL datasets, the vast search space significantly hinders retrieval efficiency. With relatively small performance gains, row-level embeddings may not be practical for large-scale table retrieval.}

\begin{figure}
\centering
        \includegraphics[width=0.48\textwidth]{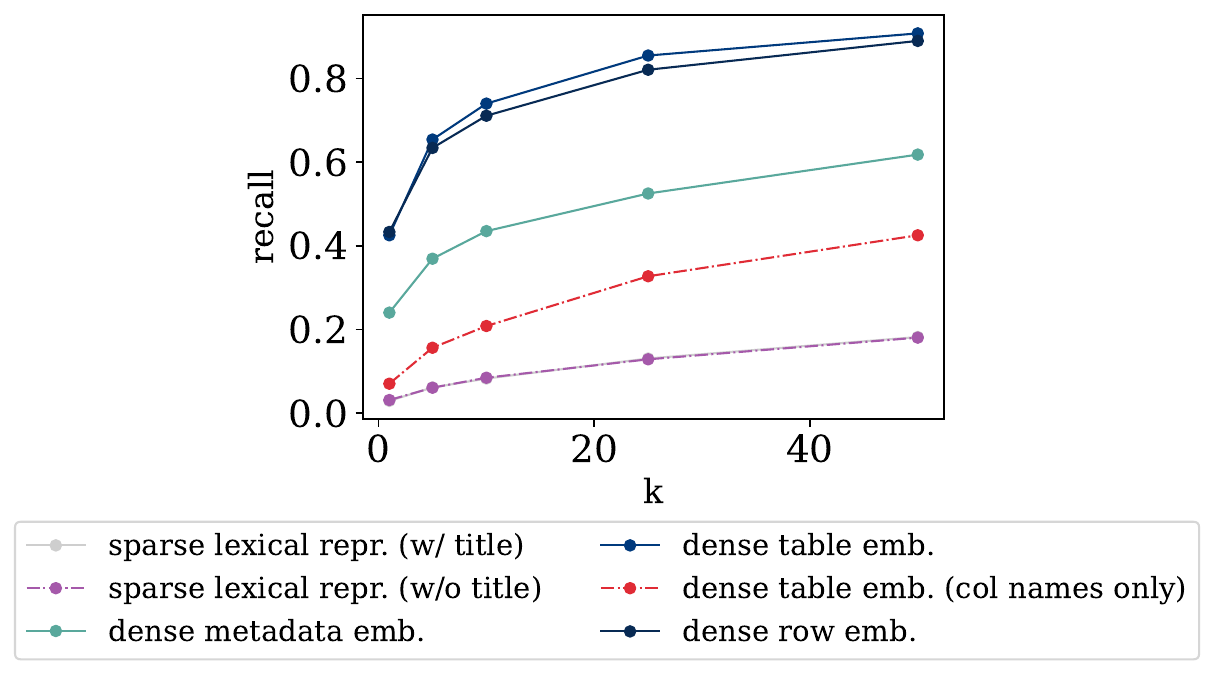}
        \caption{\textcolor{blue}{Influence of $k$ on retrieval performance with various baselines on the FeTaQA dataset, confirming the expectation that performance gradually increases with $k$, most significantly for dense embedding approaches.}}\label{fig:top-k}
\end{figure}
\begin{figure}
\centering
        \includegraphics[width=0.34\textwidth]{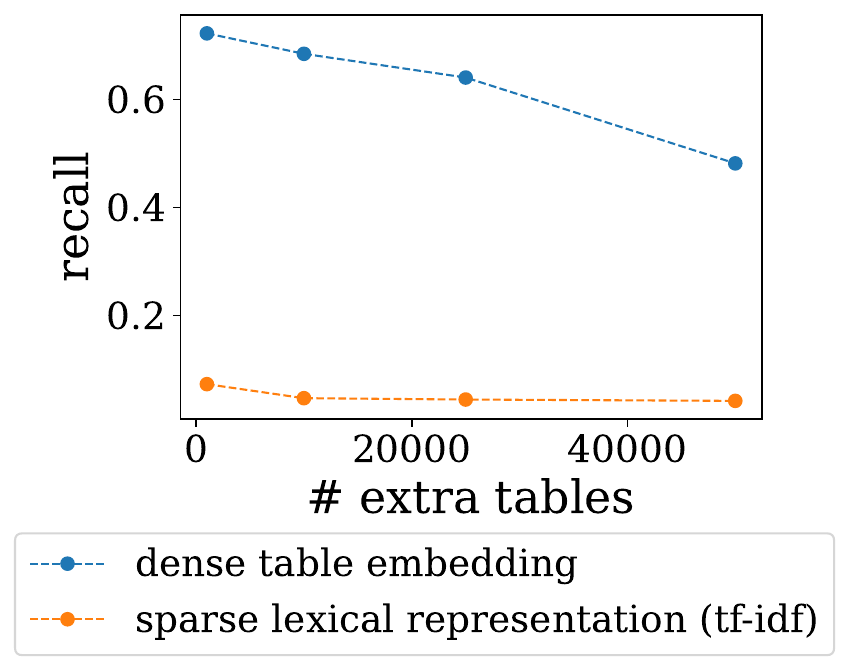}
        \caption{\textcolor{blue}{Influence of corpus size on retrieval, illustrating the sensitivity in retrieval performance of dense retrievers when the corpus reaches a large scale.}}\label{fig:nih}
\end{figure}

\paragraph{How important is table metadata for retrieval?}
From our analysis of the retrieval results of methods based on sparse lexical representations (OTTQA TF-IDF and BM25), we conclude that descriptive metadata (e.g. table summaries or titles) can be key for lexical retrievers. We observe a similar sensitivity for lexical representations for semantic metadata on the Text-to-SQL tasks when table names are not included, which is further confirmed with results on FeTaQA, where the provided table titles are not descriptive (e.g. ``example-10461'') and including them does not enhance performance. The importance of metadata is also highlighted in the strong performance of the dense metadata embedding method compared to the dense table embedding method for text-to-SQL.

\paragraph{How does scale affect retrieval performance?}

First, we assess the impact of the number of retrieved tables, i.e. by increasing $k$. As Figure~\ref{fig:top-k} shows, average recall gradually increases with $k$ for all retrievers, which is expected. The lexical retrievers do not gain significant performance improvements upon retrieving more tables.



Another influential variable is the size of the retrieval corpus. To analyze this, we evaluate the retrieval performance as corpus size increases, by appending tables from the GitTables dataset~\cite{hulsebos2023gittables}. Here we zoom in on the FeTaQA dataset, which initially consists of 2K tables. We study the impact of corpus size on retrieval performances of the sparse lexical baseline based on TF-IDF and the dense table embedding baseline. As Figure~\ref{fig:nih} shows, retrieval performance decreases  as the corpus size grows. For the dense table embedding baseline, which generally exhibits the best performance across tasks, the drop becomes progressively more noticeable once the corpus exceeds 10K added tables. Performance degradations on large corpora illustrates a need for developing table retrievers that remain robust at scale.


\subsection{Generator insights}
\begin{table*}[h]
\textcolor{blue}{
	\centering
	\caption{\textcolor{blue}{Results with {\target} for downstream tasks corresponding upfront table retrieval with $k$=10. SB stands for SacreBleu, EX for execution accuracy aggregated over all query complexity categories. P/R/F1 reflect precision, recall, and f1 scores. For Dense Table Embedding, we report the results of the best performing embedding model \texttt{stella\_en\_400M\_v5} Best scores are in \textbf{bold}, second-best \underline{underlined}.}}\label{tab:downstream-results}
    \vspace{0.1cm}
    \renewcommand{\arraystretch}{1.2}
	\resizebox{\textwidth}{!}{%
		\begin{tabular}{l : c c  : c : c c}
			\toprule
			& \multicolumn{2}{: c :}{\textbf{\small{Question Answering}}} & \textbf{Fact Verification} & \multicolumn{2}{c}{\textbf{Text-to-SQL}} \\
			& \textbf{OTTQA (SB)}  &  \textbf{FeTaQA (SB)} & \textbf{TabFact (P/R/F1)}  & \textbf{Spider (EX)} & \textbf{BIRD (EX)}\\
			\midrule
			No Context     & 0.146 & 6.761  & 0.59/0.19/0.25 & 0  & 0 \\
			Sparse Lexical Repr. (BM25) & 0.475  &  1.618  &  0.64/0.25/0.33  & 0.444 & 0.076 \\
			Sparse Lexical Repr. (TF-IDF) &  \textbf{0.510} &  1.586 &  0.64/0.25/0.33  & 0.440 & 0.183 \\
			Dense Metadata Embedding  &	0.476  &   11.027  & 0.63/0.34/0.40  &	0.556  &	\underline{0.266}	 \\
            Dense Table Embedding & \underline{0.486} & \underline{12.569}  &  \underline{0.64/0.47/0.49}  &   \underline{0.588}  &  \textbf{0.291} \\
            Dense Row-level Embedding  &  0.469 &  \textbf{13.231}  &  \textbf{0.64/0.48/0.50}  &   \textbf{0.599}  &  N/A \\
			\bottomrule
		\end{tabular}
	}
    \vspace{0.1cm}
    }
\end{table*}

\paragraph{Can LLMs execute tabular tasks from memory?} In general, the ``No Context'' baseline performs significantly lower without having relevant tables provided (Table~\ref{tab:downstream-results}). An exception to this is the low performance of sparse lexical retrievers on FeTaQA, which we discuss in the next section. Without grounding LLMs in relevant structured data to answer domain-specific questions, factuality and quality of generation becomes unreliable. Additionally, Table ~\ref{tab:downstream-results} also emphasizes that database schemas for text-to-SQL are critical to generate accurate SQL queries, as the ``No Context'' baseline yields an accuracy of 0. 

\paragraph{Does generation benefit from table retrieval?} The low performance of all retrievers on the OTTQA dataset is notable (all SacreBleu scores are below 1), which we hypothesize is due to the relatively short answers in OTTQA versus longer generated answers despite prompting for conciseness. In comparison to the ``No Context'' baseline, where the model is asked to generate answers solely based on its knowledge base, providing retrieved tables in context increases downstream performances notably, as exemplified by results for FeTaQA and TabFact. Due to the stronger retriever performance of dense embeddings, we find that dense retrievers generally yield best downstream performance across datasets. Meanwhile, the poor retrieval performance of sparse lexical representations on FeTaQA seems to distract the generator with irrelevant tables, leading to a significant decrease in SacreBleu scores compared to the ``No Context'' baseline. Ensuring the inclusion of relevant tables in the LLM’s context is crucial for reliable downstream generation quality, highlighting the need for robust retrieval methods.

\begin{figure}[h]
	\centering
	\includegraphics[width=\columnwidth]{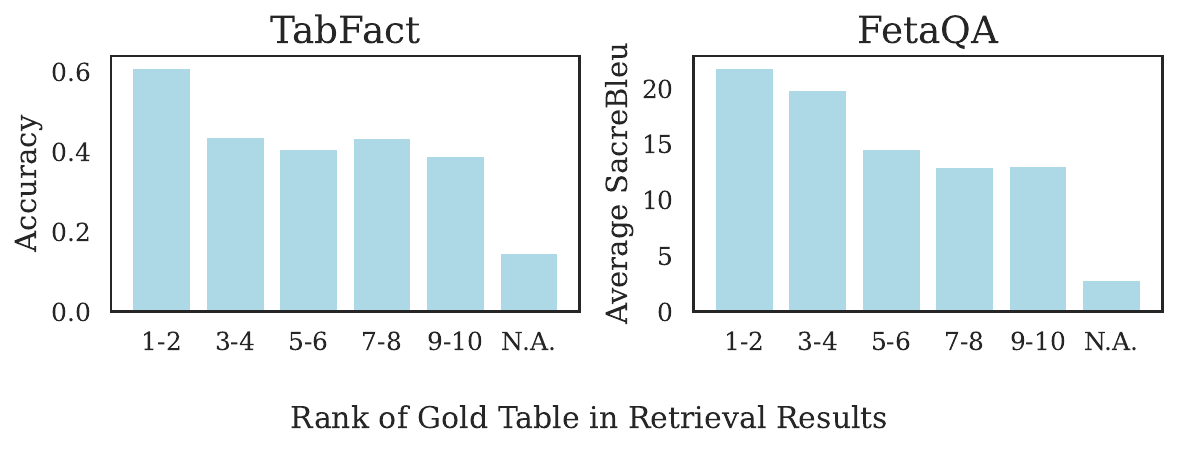}
	\caption{As the rank of the ground-truth table increases (i.e. the table is deeper in the prompt), the performance on downstream tasks tends to decrease. ``N.A.'' indicates that the ground-truth table was not retrieved.} 
	\label{fig:retriever-rank}
    \vspace{-0.2cm}
\end{figure}

\paragraph{Can long-context LLMs replace table retrieval?} An alternative for retrieval-augmented generation (RAG) is to exhaust the context of LLMs by including vast amounts of tables from the corpus without fine-grained retrieval, and rely on the LLM to extract the answer from a large set of tables. To understand the limitations of LLM context for table comprehension tasks, we explore the relationship between the rank of the ground-truth table in the retrieval results and downstream task performance in Figure \ref{fig:retriever-rank}. Treating instances where the ground-truth table failed to appear in the top-10 retrieval results as the lowest rank, we see a strong negative correlation (average Spearman's $\rho=$ -0.85) between retriever performance and downstream task performance\footnote{Due to the multi-table retrieval setting of text-to-SQL, we only consider Question Answering and Fact Verfication here.}. These results 1) motivate work on improved table retrieval and reranking, and 2) indicate that careful attention in crafting table retrievers is more effective than relying on providing a large number of tables into long-context LLMs.

%% file: sections/5_conclusion.tex
\section{Conclusion}\label{sec:conclusion}
Retrieval is key in LLM-powered query systems over structured data as well as for grounding dialog and interactions with LLMs in up-to-date, factual, structured data. With both categories of use-cases in mind, we present {\target}: the first benchmark for Table Retrieval for Generative Tasks. With {\target} we extend the ``open-domain'' query setting beyond question answering, i.e. for fact verification and text-to-SQL, to evaluate end-to-end query systems that can encompass basic retrieval and generation, as well as more complex, potentially agentic, multi-step pipelines. We evaluate various retrieval methods based on different representations of tabular data, and find that sparse lexical representations are not as robust as found for retrieval over text. Instead, dense embeddings of tables or their metadata are critical for identifying the relevant tables for given queries, with the impact of different types of metadata being an important aspect to study further. We evaluate end-to-end query performance as well, and find that table retrieval significantly improves the accuracy of LLM generations across tasks and datasets. Through deeper analysis with {\target} we surface the importance for more robust table retrievers as retrieval performance declines for large corpus sizes, whereas generation accuracy is affected by the position of the relevant table provided in the context.

\clearpage
\section{Acknowledgments}
We acknowledge support from the Dutch Research Council (NWO) grant NGF.1607.22.045, a BIDS Accenture Fellowship, the National Science Foundation (grants DGE-2243822, IIS-2129008, IIS-1940759, IIS-1940757), and EPIC lab sponsors (G-Research, Adobe, Microsoft, Google).

\section{Limitations}
{\target} does not incorporate all existing table retrieval methods due to lacking source code and tuned models, but the baselines included reflect the main method categories, varying in representation (e.g. dense or sparse) and inputs (e.g. with and without table metadata, or metadata-only). We make it straightforward to evaluate custom retrieval methods with the {\target} API, which is inspired by the MTEB benchmark for text embedding evaluation~\cite{muennighoff2023mteb}. We encourage researchers to directly integrate their retrieval methods into {\target} to establish a comprehensive and consistent ground for evaluation.

We also note that relational databases can be large, hence, necessitate more fine-grained retrieval (i.e. row and column selection~\cite{chen2024tablerag}). In {\target}, we extract ground-truth labels of the relevant database and tables for text-to-SQL datasets enabling two-step retriever evaluations (retrieving database first, then the relevant tables within that database). While ground-truth column- and row-level labels of relevance for a given query are lacking, `in-table` retrieval can be evaluated indirectly by assessing generation performance for retrieved table fragments.

Finally, upon inspecting the generation results we observed a discrepancy between the conciseness of the ground-truth answers (e.g. in OTTQA~\cite{chen2020ottqa}) and the comprehensiveness of the generated answers by GPT-4o, despite instructing for concise responses. The SacreBleu score can be sensitive to such differences. The BERTScore~\cite{bert-score} metric is commonly used for evaluating long-form QA systems as it measures semantic similarity, but our evaluations with this metric provided no discriminative signal across retrievers, as the generated answer might semantically overlap regardless of being correct with the ``pointwise'' answer, this motivates further development of suitable metrics to evaluate long-form answers.

%% file: sections/6_appendix.tex
\section*{Appendix}\label{app}

\section{Data characteristics}\label{app:data-characteristics}

\begin{figure}[h] 
    \centering
    \begin{minipage}{0.5\textwidth}
        \centering
        \begin{subfigure}{1\textwidth}
            \centering
            \begin{subfigure}{0.49\textwidth}
                \centering
                \includegraphics[width=\textwidth]{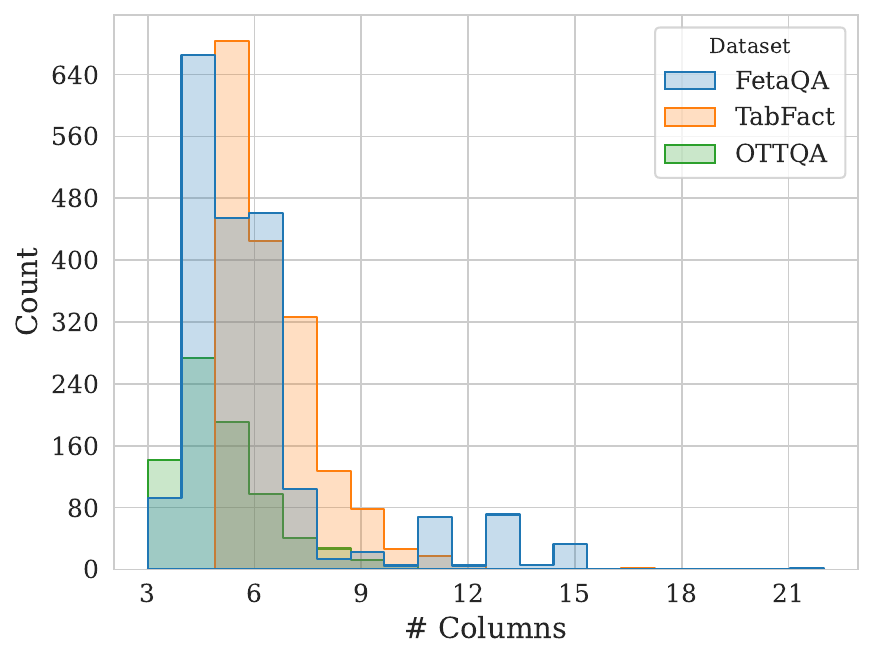}
            \end{subfigure}
            \hfill
            \begin{subfigure}{0.49\textwidth}
                \centering
                \includegraphics[width=\textwidth]{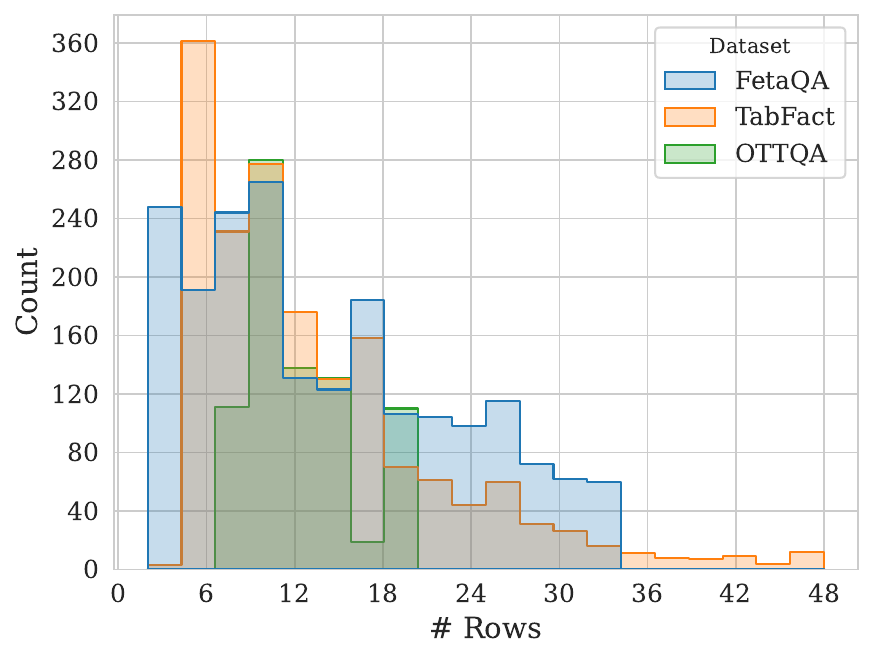}
            \end{subfigure}
            \caption{Question Answering and Fact Verification Datasets}
        \end{subfigure}
        \vspace{0.5em}
        \begin{subfigure}{1\textwidth}
            \begin{subfigure}{0.49\textwidth}
                \centering
                \includegraphics[width=\textwidth]{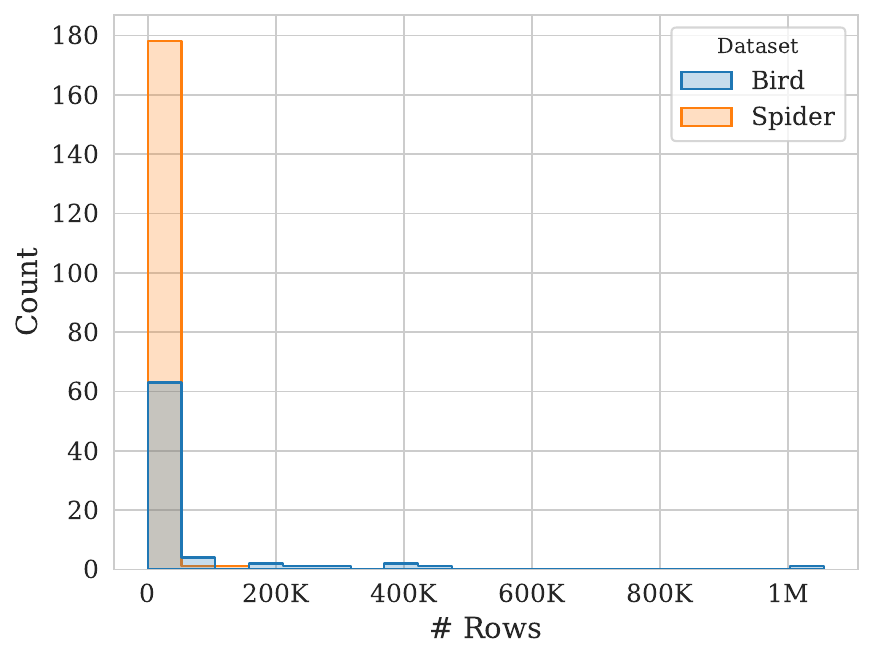}
            \end{subfigure}
            \hfill
            \begin{subfigure}{0.49\textwidth}
                \centering
                \includegraphics[width=\textwidth]{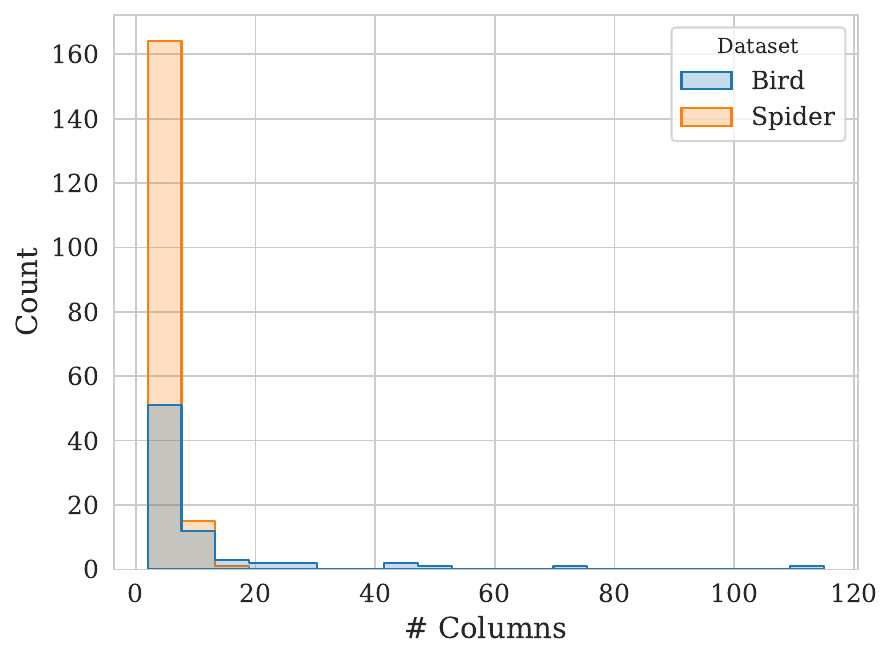}
            \end{subfigure}
            \caption{Text-to-SQL Datasets}
        \end{subfigure}
    \end{minipage}

    \caption{\textcolor{blue}{Comparison of column and row count distributions across datasets. The x-axis shows the number of columns / rows per table, and the y-axis represents the number of tables with that column / row count. Text-to-SQL datasets are separated from the Question Answering and Fact Verification datasets due to the significant differences in row count distributions.}}\label{fig:table-dimensions}
\end{figure}

\textcolor{blue}{Besides differentiation in task queries and corpus sizes, the datasets in {\target} present various distinctive properties in terms of data distributions such as table dimensions as shown in Figure \ref{fig:table-dimensions}.}

\section{Generator Prompt Templates}\label{app:prompts}

\paragraph{Question Answering} These prompts were adapted from~\cite{zhuang2024structlm}.

\noindent System Prompt:
\begin{prompt}
You are a data analyst who reads tables to answer questions.
\end{prompt}

\noindent Instruction Prompt:
\begin{prompt}
Use the provided table(s) to answer the question. Yield a concise answer to the question.\\
If none of the tables provide relevant information, use your knowledge base to generate an answer — but only if you are confident in the answer's factuality.\\
If the neither the tables nor your knowledge can be used to answer the question reliably, say that not enough information is provided.\\

Tables: {\color{blue}\{table\_contents\}}\\
Question: {\color{blue}\{query\}}
\end{prompt}

\paragraph{Fact Verification} These prompts were adapted from~\cite{zhuang2024structlm}.

\noindent System Prompt:
\begin{prompt}
You are an expert in evaluating statements on factuality given the provided tables.
\end{prompt}

\noindent Instruction Prompt:

\begin{prompt}
Given the following evidence which may take the form of sentences or a data table,determine whether the evidence supports or refutes the following statement.\\
If none of the tables provide relevant information, refer to your knowledge base — but only if you are confident your answer's factuality. If the neither the evidence nor your knowledge can be used to verify the statement reliably, state that there is not enough information.\\
Assign the statement one of three labels: True, False, Not Enough Information. Do not include anything else in your answer.\\

Tables: {\color{blue}\{table\_contents\}}\\
Statement: {\color{blue}\{query\}}
\end{prompt}

\paragraph{Text-to-SQL Prompts} These prompts were adapted from CHESS~\cite{talaei2024chess}.

To ensure the generated SQL query can be easily parsed from the generator's response (which includes both Chain of Thought Reasoning and the generated SQL), we use OpenAI's structured output API to enforce output in JSON format. 

\noindent System Prompt:

\begin{prompt}
You are an expert and very smart data analyst.
\end{prompt}

\noindent Instruction Prompt:
\begin{prompt}
Below, you are presented with a database schema and a question.\\
Your task is to read the schema, understand the question, and generate a valid SQLite query to answer the question.\\
Before generating the final SQL query, think step by step on how to write the query.\\
Database Schema: {\color{blue}\{database\_schema\}}\\
This schema offers an in-depth description of the database's architecture, detailing tables, columns, primary keys, foreign keys, and any pertinent
information regarding relationships or constraints.\\

Question: {\color{blue}\{query\}}\\

Take a deep breath and think step by step to find the correct SQLite SQL query. If you follow all the instructions and generate the correct query,
I will give you 1 million dollars.
\end{prompt}

\paragraph{No Context Instruction}
For "No Context" baseline evaluations, we provide the following message to the generator in place of the {\color{blue}\{table\_contents\}} field in the instruction prompts.

\begin{prompt}
Some or all tables are not available. Don't acknowledge the lack of information in your response. Please use your knowledge base and answer to the best of your ability, without producing false information.
\end{prompt}